\documentclass{cupconf}
\usepackage{epsfig}

  \checkfont{eurm10}
  \iffontfound
    \IfFileExists{upmath.sty}
      {\typeout{^^JFound AMS Euler Roman fonts on the system,
                   using the 'upmath' package.^^J}%
       \usepackage{upmath}}
      {\typeout{^^JFound AMS Euler Roman fonts on the system, but you
                   dont seem to have the}%
       \typeout{'upmath' package installed. cupconf.cls can take advantage
                 of these fonts,^^Jif you use 'upmath' package.^^J}%
      }
  \else
  \fi

  \checkfont{msam10}
  \iffontfound
    \IfFileExists{amssymb.sty}
      {\typeout{^^JFound AMS Symbol fonts on the system, using the
                'amssymb' package.^^J}%
       \usepackage{amssymb}%

      }{}
  \fi

  \IfFileExists{amsbsy.sty}
    {\typeout{^^JFound the 'amsbsy' package on the system, using it.^^J}%
     \usepackage{amsbsy}}
    {}

\newsavebox{\astrutbox}
\sbox{\astrutbox}{\rule[-5pt]{0pt}{20pt}}

\newcommand\etal{\mbox{\textit{et al.}}}

\title[Red-edge Spectral Signature]{The Vegetation Red Edge Spectroscopic Feature as a Surface Biomarker}

\author[Seager \& Ford]%
{S. \ns S\ls E\ls A\ls G\ls E\ls R$^{1,2}$
\and E.\ns B.\ns F\ls O\ls R\ls D$^3$}

\affiliation{$^1$Department of Terrestrial Magnetism, Carnegie Institution of Washington, Washington, DC 20015, USA\\[\affilskip]
$^2$School of Natural Sciences, Institute for Advanced Study, Princeton, NJ 08540, USA\\[\affilskip]
$^3$Princeton University Observatory, Princeton, NJ 08544, USA}

\pubyear{2003}
\volume{???}
\pagerange{??}
\date{?? and in revised form ??}
\begin{document}

\maketitle

\begin{abstract}
The search for Earth-like extrasolar planets is in part motivated by
the potential detection of spectroscopic biomarkers. Spectroscopic
biomarkers are spectral features that are either consistent
with life, indicative of
habitability, or provide clues to a planet's
habitability. Most attention so far has been given to atmospheric
biomarkers, gases such as O$_2$, O$_3$, H$_2$O, CO, and CH$_4$.  Here
we discuss surface biomarkers. Surface biomarkers that have large,
distinct, abrupt changes in their spectra may be detectable in an
extrasolar planet's spectrum at wavelengths that penetrate to the
planetary surface. Earth has such a surface biomarker: the vegetation
``red edge'' spectroscopic feature. Recent interest in Earth's surface
biomarker has motivated Earthshine observations of the spatially
unresolved Earth and two recent studies may have detected the
vegetation red edge feature in Earth's hemispherically integrated
spectrum. A photometric time series in different colors should help in
detecting unusual surface features in extrasolar Earth-like planets.
\end{abstract}

\firstsection 
\section{Introduction}
One hundred extrasolar giant planets are currently known to orbit
nearby sun-like stars.  These planets have been detected by the radial
velocity method and so, with the exception of the one transiting planet,
only the minimum mass and orbital parameters are known.  Many plans are
underway to learn more about extrasolar planets' physical properties
from ground-based and space-based observations and via proposed or
planned space missions.  Direct detection of scattered or thermally
emitted light from the planet itself is the only way to learn about a
variety of the planet's physical characteristics. Direct detection of
Earth-size planets, however, is extremely difficult because of the
proximity of a parent star that is $10^6$ to $10^{10}$ times brighter
than the planet.

Terrestrial Planet
Finder
(TPF), with a launch date in the 2015 timeframe, is being planned by
NASA to find and characterize terrestrial-like planets in the
habitable zones of nearby stars. The ESA mission
Darwin
has similar goals. The motivation for both of these space missions is
the detection and spectroscopic characterization of extrasolar
terrestrial planet atmospheres. Of special interest are atmospheric
biomarkers---such as O$_2$, O$_3$, H$_2$O, CO and CH$_4$---which are
either indicative of life as we know it, essential to life, or can
provide clues to a planet's habitability (\cite[Des Marais \etal\
2002]{DM02}).  In addition, physical characteristics such as
temperature and planetary radius could be constrained from
low-resolution spectra.

We have shown (\cite[Ford, Seager \& Turner 2001]{FST01}) that planet
characteristics could also be derived from photometric measurements of
the planet's variability at visible wavelengths. A time series of
photometric data of a spatially unresolved Earth-like planet could
reveal a wealth of information such as weather, the planet's rotation
rate, presence of large oceans or surface ice, and existence of
seasons. The amplitude variation of the time series depends on
cloud-cover fraction; more cloud cover makes a more photometrically
uniform Earth and so reduces variability. The signal-to-noise
necessary for photometric study would be obtained by a mission capable
of measuring the sought-after atmospheric biomarker spectral
features. Furthermore the photometric variability could be monitored
concurrently with a spectroscopic investigation, as was done for the
transiting extrasolar giant planet HD209458b (\cite[Charbonneau \etal\
2002]{Charb02}).

To detect and study surface properties only wavelengths that penetrate
to the planetary surface are useful. Visible wavelengths are more
suited than mid-IR wavelengths for such measurements for several
reasons. First, the albedo contrast of surface components is much
greater than the temperature variation across the planet's
surface. Second, at visible wavelengths the planet's flux is from
scattered starlight and hence at some configurations the planet is
only partially illuminated. This allows a more concentrated signal
from surface features, such as continents, as they rotate in and out of
view. Furthermore, the non-uniform illumination and non-isotropic
scattering of different surface components mean much of the scattered
light can come from a small part of the planet's surface. At mid-IR
wavelengths the planet has, to first order, uniform flux across the
planet hemisphere. In addition, the narrow transparent spectral window
at 8-12~$\mu$m will close for warmer planets than Earth and for
planets with more water vapor than Earth. However, further study at
the mid-IR ``window'' needs to be investigated.

An extremely exciting possibility, aided by a photometric time series,
is the detection of surface biomarkers in the spectrum of an
extrasolar planet. This would be possible at wavelengths that
penetrate to the planet's surface, and for surface features that
have large, distinct, abrupt changes in their spectra. Although most
surface features (e.g., ice, sand) show very little or very smooth
continuous opacity changes with wavelength, Earth has one surface
feature with a large and abrupt change: vegetation
(Figure~\ref{fig:plantspectrum}). In this paper we discuss Earth's
vegetation red-edge spectroscopic feature as a surface biomarker.

\section{The Vegetation Red Edge Spectral Feature}
All chlorophyll-producing vegetation has a very strong rise in
reflectivity at around 0.7~$\mu$m by a factor of five or more. This
red-edge spectral signature is much larger than the familiar
chlorophyll reflectivity bump at 0.5~$\mu$m, which gives vegetation
its green color.  In fact, if our eyes could see a little further to
the red, the world would be a very different place: plants would be
very red, and very bright. The glare from plants would be unbearably
high, like that of snow. The red edge is caused both by strong
chlorophyll absorption to the blue of 0.7~$\mu$m, and a high
reflectance due to plant cell structure to the red of 0.7~$\mu$m.
\begin{figure}
  \begin{center}	
  \epsfig{file=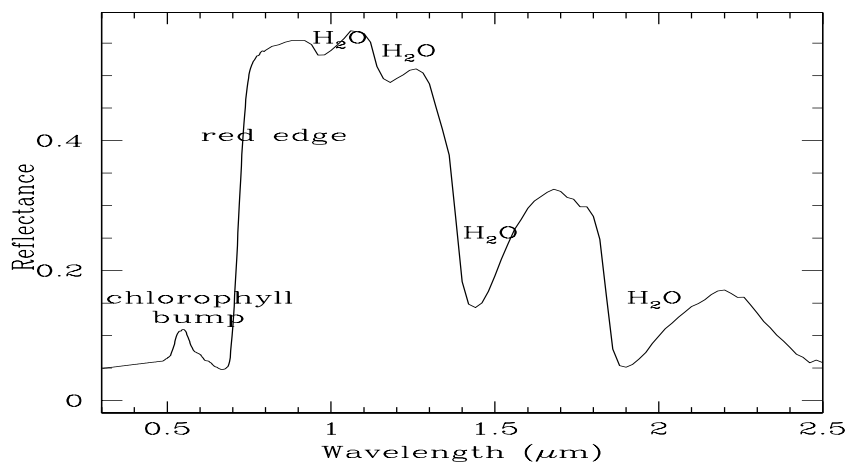,height=2.5in, width=4.5in}
  \end{center}
  \caption{Reflection spectrum of a deciduous leaf. The small bump
near 0.5~$\mu$m is a result of chlorophyll absorption (at 0.45~$\mu$m
and 0.68~$\mu$m) and gives plants their green color. The much larger
sharp rise (between 0.7 and 0.8 $\mu$m) is known as the red edge and
is due to the leaf cell structure.}
\label{fig:plantspectrum}
\end{figure}
Figure~\ref{fig:plantspectrum} shows a deciduous plant leaf reflection
spectrum. The high absorptance at UV wavlengths (not shown) and at
visible wavelengths is by chlorophyll and is used by the leaf for
photosynthesis. Photosynthesis is the process by which vegetation and some
other organisms use energy from the sun to convert H$_2$O and CO$_2$
into sugars and O$_2$. The primary molecules that absorb the light
energy and convert it into a form that can drive this reaction are
chlorophyll A (0.450~$\mu$m) and B (0.680~$\mu$m).

As seen in Figure~\ref{fig:plantspectrum}, between 0.7~$\mu$m and
1~$\mu$m the leaf is strongly reflective. Not shown in
Figure~\ref{fig:plantspectrum} is that the leaf also has a very high
transmittance at these same wavelengths, such that reflectivity plus
transparency is near 100\%. Interestingly, the bulk of the energy of
solar radiation as it reaches sea level is at approximately 0.6 to
1.1~$\mu$m.  If plants absorbed with the same efficiency at these
wavelengths as at visible wavelengths they would become too warm and
their chlorophyll would degrade.  A specific plant must balance the
competing requirements of absorption of sunlight at wavelengths
appropriate for photosynthesis reactions with efficient reflectance at
other wavelengths to avoid overheating (\cite[Gates \etal\
1965]{Gates65}).  Therefore the exact wavelength and strength of the
red edge depends on the plant species and environment.  Although
negligible from the TPF view point, it is interesting to note that the
specific wavelength and strength of the red edge feature is used for
remote sensing of specific locations on Earth to identify plant
species and also to monitor a field of vegetation's (such as crops)
health and growth as the red edge changes during the growing season.

In the near-infrared, as shown in Figure~\ref{fig:plantspectrum},
plants have water absorption bands. The band strength depends on plant
water content, weather conditions, plant type, and geographical
region.  These absorption features can be quite strong, but are not
very useful for identifying life, since they would only be indicative
of water and would not be distinguishable from atmospheric water
vapor.

Plant leaves are very reflective away from chlorophyll absorption and
water absorption wavelengths due to the internal leaf structure
(\cite[Gates \etal\ 1965]{Gates65}). Light partially scatters off of
the leaf surface but also scatters efficiently inside the leaf.  Light
reflects off of and refracts through cell walls from the surrounding
air gaps between cells.  Inside cells themselves the high change in
the index of refraction from 1.33 for water to 1.00 for air causes an
efficient internal reflection at the interface between cell walls and
the surrounding air gaps. Also, inside cells light can Mie or Rayleigh
scatter off of cell organelles which have sizes on the order of the
wavelength of light.  The overall reflectance and transmittance is a
complex function of the cell size and shape and the size and shape of
the air gaps between the cells (see, e.g., \cite[Govaerts \etal\
1996]{Gv96}).  Because there is little absorption away from the
chlorophyll and water absorption wavelength regions, light will
eventually scatter out of the leaf at the top (reflection) or bottom
(transmission).

\section{Plants as an Earth Surface Biomarker}

The red-edge spectroscopic feature is very strong for an individual
plant leaf, at a factor of five or more. Averaged over a (spatially
unresolved) hemisphere of Earth, however, the vegetation red-edge
spectral feature is reduced from this high reflectivity down to a few
percent. This is because of several effects including the forest
canopy architecture, soil characteristics, the non-continuous coverage
of vegetation across Earth's surface, and the presence of clouds which
prevent viewing the surface. In addition the reflectance of vegetation
is anisotropic and so the illumination conditions and viewing angle
are important. Nevertheless at a signal of a few percent Earth's
vegetation red edge may be a viable surface biomarker to a distant,
telescope-bearing civilization. The chlorophyll bump at 0.5~$\mu$m,
however, is negligible in a hemispherically averaged spectrum. The
spectral signature of oceanic vegetation or plankton is also unlikely
to be detectable, due to strong absorption by particles in the water
and also by the strong absorptive nature of liquid water beyond red
wavelengths.

Using vegetation's red edge as a surface biomarker is not a new idea.
Early last century the high near-infrared reflection signature was
used to test the hypothesis that the changing dark patches on Mars
were due to seasonal changes of vegetation (\cite[Slipher
1924]{Slipher24}; \cite[Millman 1939]{Millman39}; \cite[Tickhov
1947]{Tickhov47}; \cite[Kuiper 1949]{Kuiper49}). Not surprisingly,
only negative results were obtained.

More recently \cite[Sagan \etal\ (1993)]{Sagan93} used the Galileo
spacecraft for a ``control experiment'' to search for life on Earth
using only conclusions derived from data and first principle
assumptions. En route to Jupiter, the Galileo spacecraft used two
gravitational assists at Earth (and one at Venus). During the December
1990 fly-by of Earth, the Galileo spacecraft took low-resolution
spectra of different areas of Earth. In addition to finding ``abundant
gaseous oxygen and atmospheric methane in extreme thermodynamic
disequilibrium'', \cite[Sagan \etal\ (1993)]{Sagan93} found ``a widely
distributed surface pigment with a sharp absorption edge in the red
part of the visible spectrum'' that ``is inconsistent with all likely
rock and soil types''. Observing $\sim$100 km$^2$ areas of Earth's
surface the vegetation red edge feature showed up as a reflectance
increase of a factor of 2.5 between a band centered at 0.67~$\mu$m and
one at 0.76~$\mu$m. In contrast there was no red-edge signature from
non-vegetated areas.

\begin{figure}
  \begin{center}	
  \epsfig{file=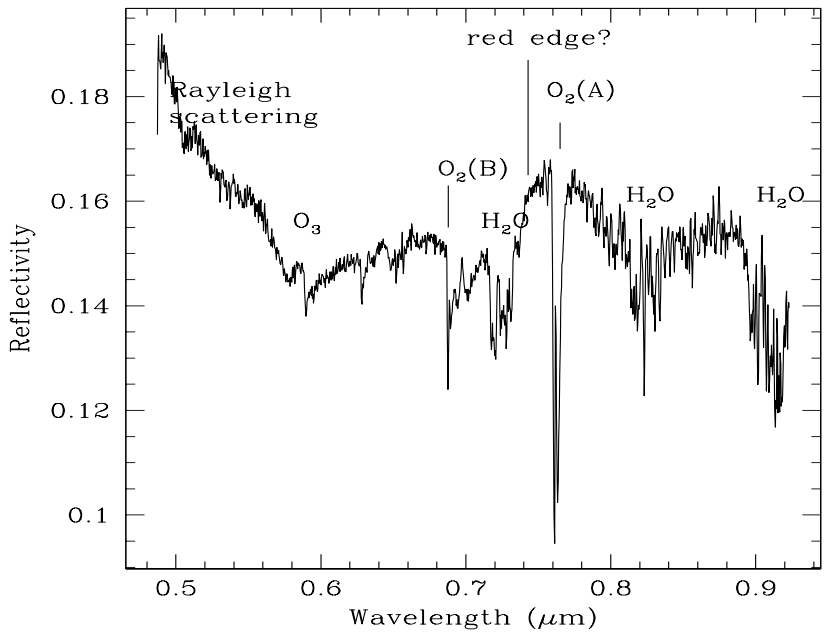,height=3.5in, width=4.5in}
  \end{center}
  \caption{ A visible wavelength spectrum of the spatially
 unresolved Earth, as seen with Earthshine 
(adapted from \cite[Woolf \etal\ 2002]{Woolf02}). The viewpoint is
largely centered equatorially on the Pacific ocean.  The major
atmospheric features are identified. The reflectivity scale is
arbitrary. Data courtesy of N. Woolf and W. Traub. For
details see Woolf \etal\ 2002.}
\label{fig:Earthshine}
\end{figure}

A new area of extrasolar planet research is now emerging: using
Earthshine to study the spatially unresolved Earth. Earthshine is
light from the sun that has been scattered off of Earth onto the moon
and then back to Earth. It appears as a faint glow on the otherwise
dark part of the moon during the crescent phase, but can be studied
with a CCD camera and specialized coronagraph even as the moon waxes
(\cite[Goode \etal\ 2001]{Goode01}). Satellite data of Earth is not as
useful as Earthshine because it is highly spatially resolved and
limited to narrow spectral regions. Also, since most satellite data is
collected by looking straight down at specific regions of Earth
hemispherical flux integration with lines-of-sight through different
atmospheric path lengths is not available. Recent spectral
observations of Earthshine have tentatively detected the red-edge
signature at the few percent level.  \cite[Woolf \etal\
(2002)]{Woolf02} observed the setting crescent moon from Arizona which
corresponds to Earth as viewed over the Pacific Ocean. Nevertheless
their spectrum (Figure~\ref{fig:Earthshine}) shows a tantalizing rise
just redward of 0.7~$\mu$m that is tentatively the spectroscopic
red-edge feature. Figure~\ref{fig:Earthshine} also shows other
interesting features of Earth's visible-wavelength spectrum, notably
O$_2$ and H$_2$O absorption bands (note that that spectral lines of
both O$_2$ and H$_2$O cut into the red-edge signature.)  \cite[Arnold
\etal\ (2002)]{Arnold02} have made observations of Earthshine on
several different dates. With observations from France the Earthshine
is from America and the Pacific Ocean (the evening moon) and Europe
and Asia (the morning moon). After subtracting Earth's spectrum to
remove the contaminating atmospheric absorption bands they find a
vegetation red edge signal of 4 to 10 \%.

\section{Temporal Variability to Detect Surface Biomarkers}
A small but sharp spectral feature from a component of the planet's
surface should be more easily identified by temporal variation. As the
continents rotate in and out of view, the planet's reflectivity will
change, causing a change in the measured spectrum.  Recent Earthshine
measurements have shown that detection of Earth's vegetation-red edge
is tricky due to smearing out by other atmospheric and surface
features.  Trying to identify such small features at unknown
wavelengths in an extrasolar planet spectrum may be very difficult.
We propose that such spectral features could be much more easily
identified by the increased temporal variabiltiy at a carefully chosen
color.  In particular, any changes associated with a rotational period
would be highly relevant.  Since the wavelength of any surface
biomarkers would not be known apriori, flexible data acquistion is
essential. For example low-resolution spectra could be later
integrated into narrow-band photometry of many different bands. Here
we discuss simulations of Earth's temporal variability, including
preliminary calculations of Earth's vegetation red-edge variability.

We model the photometric flux from a rotating Earth by a Monte Carlo
code using a spherical map of Earth which specifies the scattering
surface type at each point on the sphere and a set of
wavelength-dependent bidirectional reflectance distribution functions
which specify the probability of light incident from one direction to
scatter into another direction for each type of scattering surface
(see \cite[Ford \etal\ 2001]{Ford01} for details).  We use a map of
Earth from a one-square-degree satellite surface map that classifies
each pixel as permanent ice, dirty/temporary ice, ocean, forest,
brush, or desert.  We consider cloudy models using the scattering
properties of Earth clouds and we also include an approximation of
atmospheric Rayleigh scattering.  We focus our attention to quadrature
(a phase angle of 90$^{\circ}$) for which the planet-star separation
is largest and the observational constraints thus least severe.

\begin{figure}
  \begin{center}	
  \epsfig{file=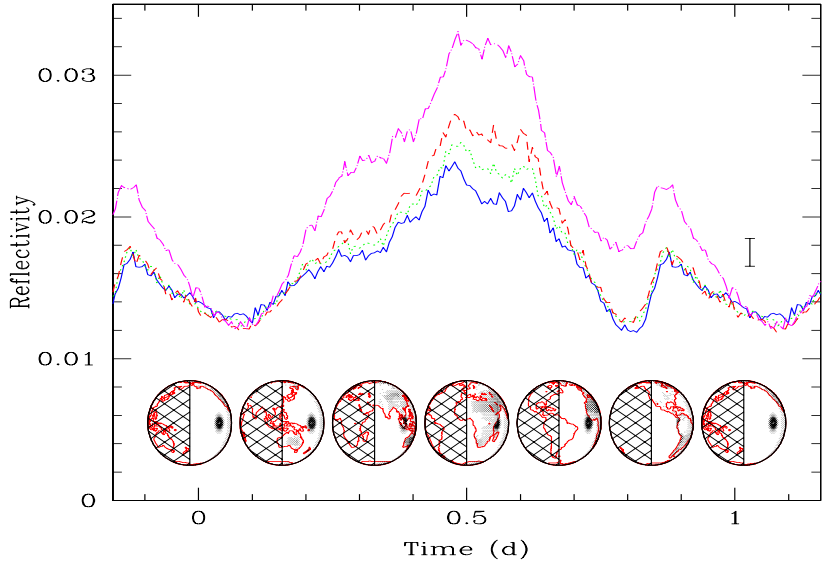,height=3.1in, width=4.5in}
  \end{center}
  \caption{ 
A light curve for a cloud-free Earth model for one rotation.  The
$x$-axis is time and the $y$-axis is the reflectivity normalized to a
Lambert disk at a phase angle of 0$^{\circ}$. The viewing geometry is
shown by the Earth symbols, and a phase angle of 90$^{\circ}$ is
used. Note that a different phase angle will affect the reflectivity
due to a larger or smaller fraction of the disk being illuminated;
because of the normalization the total reflectivity is $\ll$ in this
case of phase angle of 90$^{\circ}$. From top to bottom the curves
correspond to wavelengths of 0.75, 0.65, 0.55, and 0.45~$\mu$m, and their
differences reflect the wavelength-dependent albedo of different
surface components. The noise in the light curve is due to Monte Carlo
statistics in our calculations. The images below the light curve show
the viewing geometry (cross-hatched region is not illuminated) and
relative contributions from different parts of the disk (shading
ranges from $<$ 3\% to $>$ 40\%, from white to black) superimposed on
a map of the Earth. At time = 0.5 day, the Sahara desert is in view and
causes a large peak in the light curve due to the reflectivity of sand
which is especially high in the near-infrared (top curve).}
\label{fig:earth}
\end{figure}

\begin{figure}[b]
  \begin{center}	
  \epsfig{file=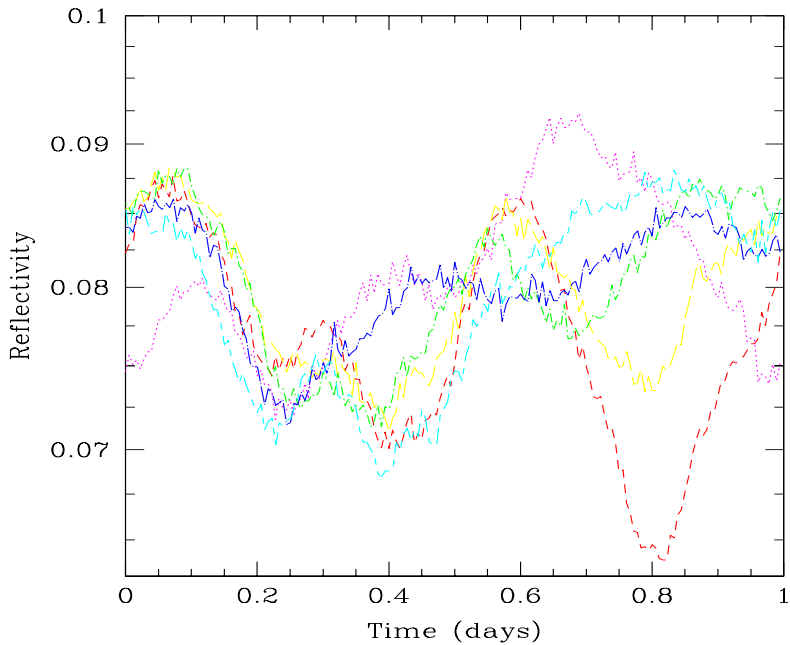,height=3.5in, width=4.5in}
  \end{center}
  \caption{ Rotational light curves for model Earth with clouds.  
This figure shows
six different daily light curves at 0.55~$\mu$m for our Earth model with
clouds, as viewed from a phase angle of 90$^{\circ}$.  These theoretical
light curves use cloud cover data from satellite measurements taken on six
consecutive days.}
\label{fig:cloudy}
\end{figure}

The existence of different surface features on a planet may be
discernable at visible wavelengths as different surface features
rotate in and out of view. Considering a cloud- free Earth, the
diurnal flux variation caused by different surface features rotating
in and out of view could be as high as 200\%
(Figure~\ref{fig:earth}). This high flux variation is not only due to
the high contrast in different surface components' albedos, but also
to the fact that a relatively small part of the visible hemisphere
dominates the total flux from a spatially unresolved planet. Clouds
interfere with surface visibility and in the presence of clouds the
diurnal light curve shown in Figure~\ref{fig:earth} becomes that shown
in Figure~\ref{fig:cloudy}. It is very interesting to note that an extrasolar
Earth-like planet certainly could have a lower cloud cover fraction
than Earth's 50\% cloud cover. The cloud pattern and cover fraction
are influenced by a variety of factors including the planet's rotation
rate, continental arrangement, obliquity, and presence of large bodies
of water.


\begin{figure}
  \begin{center}	
  \epsfig{file=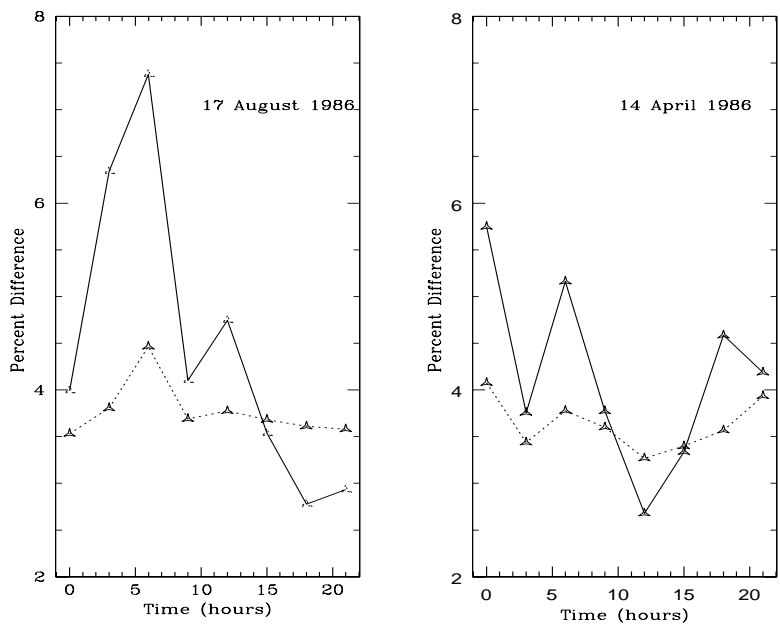,height=3.5in, width=4.5in}
  \end{center}
  \caption{ Variability of Earth's color. The solid line shows a color
[(I(0.75-0.8)-I(0.7- 0.65))/I(0.75-0.8)] chosen to emphasize
variability of vegetation's red edge. For comparison, the dotted line
shows a color [I(0.85-0.8)-I(0.75-0.8)/I(0.75-0.8)], which is less
sensitive to vegetation. These colors include theoretical spectra from
\cite[Des Marais \etal\ 2002]{DesMarais02} modulated by our Earth rotational surface and cloud model. The
cloud cover for the model in the left panel is from the ICSSP database
from 17 August 1986 and in the right panel from 14 April 1986. Earth
is more variable in the color sensitive to the red edge vegetation
feature.}
\label{fig:time}
\end{figure}

A time series of data in different colors (Figure~\ref{fig:time}) may
help make it possible to detect a small but unusual spectral feature,
even with variable atmospheric features. Most of Earth's surface
features, such as sand or ice, have a continuous increase or minimal
change with wavelength, in constrast to the abrupt vegetation red edge
spectral feature. We have generated spectrophotometric variability of
Earth by using theoretical spectra (for a cloudy and non-cloudy
atmosphere from Traub (private communication) as included in
(\cite[Des Marais \etal\ 2002]{DM02})) modulated by our Earth
rotational surface and cloud model. We use cloud data from the ISCCP
database (\cite[Rossow \& Schiffer 1991]{Rossow91}) such that the
rotating Earth also has changing cloud patterns. We have chosen to
integrate the spectrum into colors, the first
[(I(0.75-0.8)-I(0.7-0.65))/I(0.75- 0.8)] chosen to emphasize
variability of vegetation's red edge and the second, for comparison, a
color [I(0.85-0.8)-I(0.75-0.8)/I(0.75-0.8)], which is less sensitive
to vegetation. Figure~\ref{fig:time} shows that Earth is more variable
in a color across the red edge than for colors with similar wavelength
differences in other parts of Earth's spectrum.  For extrasolar planet
measurements spectra or spectrophotometric data would be most useful
in the form of a spectrum so that the photometric bands can be chosen
after data acquisition.

\section{Extrasolar Plants?}
It is difficult to speculate on extrasolar plants and we will not do
so here.  Some might argue that the vegetation red edge differences
among coniferous, deciduous, and desert plants are
meaningful. However, all Earth vegetation has almost certainly evolved
from the same ancestor and it is not a fair evolutionary experiment.
Nevertheless a few interesting facts are suggestive and useful to
those who wish to speculate on the possible existence extrasolar
plants or light harvesting organisms:

$\bullet$ Plants absorb very efficiently throughout the UV
and the visible wavelength regions of the spectrum where the energy is
required for photosynthesis (involving molecular electronic
transitions);

$\bullet$ At sea level (after atmospheric extinction) the solar energy
distribution peaks at 1 $\mu$m and approximately 50\% of the energy is
redward of 0.7 $\mu$m;

$\bullet$ Plants reflect and transmit almost 100\% of light in the
wavelength region where the direct sunlight incident on plants has the
bulk of its energy;

$\bullet$ Considering these last three points, Earth's
primary ``light harvesting organism'', vegetation, has
evolved to balance the competing requirements of absorption
of sunlight at wavelengths appropriate for photosynthesis reactions
with efficient reflectance at other wavelengths to avoid overheating
(\cite[Gates \etal\ 1965]{Gates65});

$\bullet$ Other pigments involved in vegetation's light harvesting
process also absorb in the 0.44~$\mu$m wavelength regime (but only
chlorophyll B absorbs near 0.68~$\mu$m). However, some other organisms
have pigments that absorb at other wavelengths. For example,
the light-harvesting pigments carotenoids and phycobilins are
used by red algae.

\section{Summary and Conclusions}
The vegetation red edge spectroscopic feature is a factor of 5 or more
change in reflection at $\sim$ 0.7$\mu$m. This red edge feature is
well-used in satellite remote sensing studies of Earth's
vegetation. Earthshine observations have been used to detect the
vegetation red edge signature in the spatially unresolved spectrum of
Earth where it appears at the few percent level.

When discovered, observations of extrasolar Earth-like planets at
wavelenghts that penetrate to the planet's surface will be very
useful, especially for planets with much lower cloud cover than
Earth's 50\%.  A time series of spectra or broad-band photometry could
reveal surface features of a spatially unresolved planet, including
surface biomarkers. Earth's hemispherically integrated vegetation
red-edge signature is weak (a few to ten percent), but Earth-like
planets with different rotation rates, obliquities, land-ocean
fraction, and continental arrangement may well have lower cloud-cover.

While it is near impossible to speculate on spectral features of light
harvesting organisms on extrasolar planets, flexible data acquistion
will maximize scientific return. The detection of an unusual spectral
signature that is inconsistent with any known atomic, molecular, or
mineralogical signature would be fantastic. Combinations of unusual
spectral features together with strong disequilibrium chemistry would
be even more intriguing and would certainly motivate additional
studies to better understand the propsects for such a planet to harbor
life.

\begin{acknowledgments}

We thank the conference organizers for a very interesting meeting.  We
thank Ed Turner and other members of the Princeton Terrestrial Planet
Finder team for many useful discussions and Wes Traub and Ken Jucks
for use of their theoretical spectra. S. S. is supported by the
W. M.  Keck Foundation and the Carnegie Institution of Washington.
\end{acknowledgments}


\begin{thebibliography}{}


  \bibitem[Arnold \etal\ 2002]{Arnold02} \textsc{Arnold, L., Gillet,
     S., Lardiere, O., Riaud, P. \& Schneider, J.} 2002 {A test for the
     search for life on extrasolar planets. Looking for the
     terrestrial vegetation signature in the Earthshine spectrum}
     \textit{A\&A} \textbf{392}, 231--237.

 \bibitem[Charbonneau \etal\ (2002)]{Charb02}
     \textsc{Charbonneau, D. Brown, T. M., Noyes, R. W. \&  Gilliland, R.  L.}
     2002 {Detection of an extrasolar planet atmosphere}
     \textit{ApJ} \textbf{568}, 377--384.	

  \bibitem[Desmarais \etal\ (2002)]{DM02}
     \textsc{Des Marais, D. J., Harwit, M. O., Jucks, K. W., Kasting, J. F., 
 Lin, D. N. C., Lunine, J. I., Schneider, J., Seager, S., Traub, W. A., \&
  Woolf, N. J.} 2002
     {Remote sensing of planetary properties and biosignatures on
     extrasolar terrestrial planets}
     \textit{Astrobiology} \textbf{2}, 153--181.

 \bibitem[Ford, Seager \& Turner (2001)]{FST01}
     \textsc{Ford, E. B., Seager, S. \& Turner E. L.} 2001
     {Characterization of extrasolar terrestrial planets from 
      diurnal photometric variability}
      \textit{Nature} \textbf{412}, 885--887.

  \bibitem[Gates \etal\ 1965]{Gates65}
     \textsc{Gates, D. M., Keegan, H. J., Schleter, J. C. \& Weidner, V. R.}
     1965 {Spectral properties of plants}
     \textit{Applied Optics} \textbf{4}, 11--20.

  \bibitem[Goode \etal\ (2001)]{Goode01}
     \textsc{Goode, P. R., Qiu, J., Yurchyshyn, V., Hickey, J.,
      Chu, M.-C., Kolbe, E., Brown, C. T. \& Koonin, S. E.} 2001
      {Earthshine observations of the Earth's reflectance}
      \textit{GeoRL} \textbf{28}, 1671.

  \bibitem[Govaerts \etal\ 1996]{Gv96}
     \textsc{Govaerts, Y. M., Jacquemoud, S, Verstraete, M. M. \& 
     Ustin, S. L.} 1996
     {Three-dimensional radiation transfer modeling in a dicotyledon leaf}
     \textit{Applied Optics} \textbf{35}, 6585--6598.

  \bibitem[Kuiper 1949]{Kuiper49}
     \textsc{Kuiper G. P.} 1949 {The Atmospheres of the Earth and Planets}
     (Chicago: University of Chicago Press), p.339.


  \bibitem[Millman 1939]{Millman39}
     \textsc{Millman, P. M.}
     1939 {Is there vegetation on Mars?}
     \textit{The Sky} \textbf{3}, no. 10, 11.

  \bibitem[Rossow \& Schiffer 1991]{Rossow91}
     \textsc{Rossow W. B. \& Schiffer R. A.} 1991 
     {ISCCP cloud data products}
     \textit{Bull. Amer. Meteor. Soc.} \textbf{72}, 2--20.


  \bibitem[Slipher 1924]{Slipher24}
     \textsc{Slipher J. M.}
     1924 {Observations of Mars in 1924 made at the Lowell Observatory: II. spectrum observations of Mars
}     \textit{PASP} \textbf{36}, 261--262.

  \bibitem[Tikhov 1947]{Tikhov47}
     \textsc{Tikhov, G. A.}
     1947 {}  \textit{Bull. Astr. and Geodet. Soc. U.S.S.R.}
     \textbf{8}, 8.


  \bibitem[Woolf \etal\ 2002]{Woolf02} \textsc{Woolf, N. J., Smith,
     P. S., Traub, W. A. \& Jucks, K. W.}  2002 {The spectrum of
     Earthshine: a pale blue dot observed from the ground}
     \textit{ApJ} \textbf{574}, 430--433.

\end{thebibliography}
\end{document}